\def\year{2022}\relax
\documentclass[letterpaper]{article} 
\usepackage{aaai22}  
\usepackage{times}  
\usepackage{helvet}  
\usepackage{courier}  
\usepackage[hyphens]{url}  
\usepackage{graphicx} 
\usepackage{natbib}  
\usepackage{caption} 
\DeclareCaptionStyle{ruled}{labelfont=normalfont,labelsep=colon,strut=off} 
\usepackage{algorithm}
\usepackage{algorithmic}
\usepackage{booktabs}
\usepackage{hyperref}

\usepackage{newfloat}
\usepackage{listings}
\floatstyle{ruled}
\newfloat{listing}{tb}{lst}{}
\floatname{listing}{Listing}
\title{EDTok: A Dataset for Eating Disorder Content on TikTok}
\author{
Charles Bickham\textsuperscript{\rm 1},
Bryan Ramirez-Gonzalez\textsuperscript{\rm 1},
Minh Duc Chu\textsuperscript{\rm 1,\rm2},
Kristina Lerman\textsuperscript{\rm 1,\rm2},
Emilio Ferrara\textsuperscript{\rm 1}
}
\affiliations{ 
\textsuperscript{\rm 1} Department of Computer Science, Viterbi School of Engineering, University of Southern California, Los Angeles, CA, USA

\textsuperscript{\rm 2}Information Sciences Institute, University of Southern California, Marina Del Rey, CA, USA\\
cbickham@usc.edu, bryanram@usc.edu, mhchu@usc.edu, lerman@isi.edu, emiliofe@usc.edu

,
}
\usepackage{bibentry}

\begin{document}

\maketitle

\begin{abstract}
Eating disorders, which include anorexia nervosa and bulimia nervosa, have been exacerbated by the COVID-19 pandemic, with increased diagnoses linked to heightened exposure to idealized body images online. TikTok, a platform with over a billion predominantly adolescent users, has become a key space where eating disorder content is shared, raising concerns about its impact on vulnerable populations. In response, we present a curated dataset of 43,040 TikTok videos, collected using keywords and hashtags related to eating disorders. Spanning from January 2019 to June 2024, this dataset, offers a comprehensive view of eating disorder-related content on TikTok. Our dataset has the potential to address significant research gaps, enabling analysis of content spread and moderation, user engagement, and the pandemic's influence on eating disorder trends. This work aims to inform strategies for mitigating risks associated with harmful content, contributing valuable insights to the study of digital health and social media's role in shaping mental health.
\end{abstract}

\section{Introduction}
Eating disorders represent a complex mental health
condition characterized by pathologies in eating and body image. Common eating disorders include anorexia nervosa and bulimia nervosa \cite{polivy2002causes}. Studies have shown that suicide is a significant contributor to the high mortality rates among patients with anorexia nervosa \cite{zipfel2000long, herzog2000mortality, sullivan1995mortality}, with statistics indicating that one-third of individuals with anorexia nervosa and bulimia nervosa have attempted suicide \cite{smith2018eating}. COVID-19 has made the landscape of eating disorders more challenging \cite{j2023impact, schlegl2020eating}. Research shows that during this period there was an increase in symptomology and diagnoses \cite{taquet2022incidence, tavolacci2021sharp, reed2022rise, monteleone2021eating, zipfel2022hidden}. One of the main reasons for this is the increased exposure to eating disorder or anxiety provoking media \cite{rodgers2020impact}.  Social media platforms, particularly TikTok, have played a significant role in shaping these experiences.

TikTok is a user generated short video sharing app that has over a billion users worldwide \cite{iqbal2021tiktok}. It is especially popular among adolescents, a group that is already vulnerable to developing eating disorders \cite{montag2021psychology, mclean2022impacts}. It has become a space for individuals to discuss and share content related to eating disorders, including but not limited to personal experiences, recovery stories, and sometimes harmful behaviors \cite{herrick2021just, mccashin2023using, pruccoli2022use, rando2023health}. Since the beginning of the pandemic, there has been a noticeable increase in the volume of eating disorder content shared on TikTok \cite{jordan2021tiktok}, raising concerns about the platform's impact on vulnerable users. In response, TikTok has launched initiatives to moderate content that may be perceived as harmful. According to its community guidelines 
\footnote{\url{https://www.tiktok.com/community-guidelines/en/mental-behavioral-health/\#2}\label{note1}} the platform does not permit content that promotes disordered eating, dangerous weight loss behaviors, or the marketing of weight loss or muscle gain products.
However, users evade moderation by using opaque jargon and intentional misspellings — for example, writing \#edr3c0very instead of \#edrecovery.~\cite{bickham2024hidden}.

To better understand the dynamics of eating disorder content on TikTok and in response to the urgent need for research in this area, we have curated a dataset that aims to address existing research gaps. The dataset consists of over 43,040 videos, collected using a set of carefully selected eating disorder-related keywords and hashtags. This dataset offers a unique opportunity to explore the intersection of eating disorders and TikTok by providing a public, accessible resource. Unlike traditional datasets that often rely solely on self-reported data or text-based content, our dataset includes both video and text, allowing for more comprehensive analysis of how eating disorders are represented and discussed on social media.

As part of this research, we also contribute a framework for collecting metadata from the TikTok API, which facilitates the extraction of essential information and allows for the download of videos. This framework addresses significant gaps in existing research tools by enabling researchers to systematically gather and analyze TikTok data.


This dataset opens up several avenues for research, enabling scholars to address key questions about how eating disorder content spreads on TikTok, how users engage with and react to this content, and the role that social media plays in shaping perceptions of eating disorders. The data can also be used to examine the impact of the COVID-19 pandemic on the prevalence and nature of eating disorder content, offering insights into how significant global events influence mental health discussions in online spaces. Ultimately, this dataset contributes to filling a critical gap in existing research by providing a valuable resource for studying eating disorders in the context of social media, with the potential to inform future strategies for mitigating the risks associated with exposure to harmful content online.

\section{Data Collection}
We queried the TikTok Research API using a combination of previously used eating disorder hashtags \cite{bickham2024hidden, lerman2023radicalized} and a manually curated set of keywords and hashtags (see Table \ref{table:keywords}). The videos in our dataset span from January 1, 2019, to June 28, 2024, covering periods before, during, and after the COVID-19 pandemic. The metadata collected for each video includes: a unique identifier, publication time, country code, the author's username 
the video description, the unique identifier of any music used, the number of likes, comments, shares, and views, the unique identifiers of any effects applied, associated hashtags, the unique identifier of its playlist (if any), and the audio transcripts (if available).

\begin{table}[htbp!]
\caption{Keywords and Hashtags used to collect video metadata}
\centering
\begin{tabular}{@{}lll@{}}
    \toprule
    \multicolumn{3}{c}{\textbf{Keywords and Hashtags}}           \\ \midrule
    edrecovery        & edjourney              & edrec0very         \\
    ednotsheeran      & anarecoveryy           & edrecoverypositivity \\
    fearfood          & edrecoveryy            & fearfoodchallenge  \\
    edrecocery        & disorderedeatingtw     & edtiktoklive       \\
    edawareness       & edtk                   & disordereddieting  \\
    edtktk            & ednotsheereen          & edtiktok           \\
    ednosheeran       & edtww                  & ednotsheerantok    \\
    edhumor           & edproblems             & ed                \\
    miatok            & eatingdisorder         & edsheeranrecoveryy \\
    disorderedeatingtw & edtt                   & disorderedeating   \\
    anorexia          & atypicalanaorexia      & atypicalana        \\
    anarecovery       & anarecovry             & anorexiarecoveryy  \\
    anarecovr         &                        &                   \\
    \bottomrule
\end{tabular}
\label{table:keywords}
\end{table}
After collecting the metadata, we used the PykTok\footnote{\url{https://github.com/dfreelon/pyktok/}\label{note2}} Python module to download the videos in mp4 format. In total, we downloaded 72,622 videos through the TikTok API. However, PykTok failed to download posts containing photo slideshows and occasionally encountered connectivity issues, which were typically resolved by re-querying the URL after a 100-second interval. In some cases, videos were downloaded in an invalid format for analysis. After accounting for these issues, the dataset was reduced to 56,472 videos.

\section{Data Filtering}
Post collection data filtering  involved two steps. First, we randomly sampled over 500 videos to determine their relevance to eating disorders. Approximately 20\% of the sampled videos were unrelated. We manually reviewed the metadata of these unrelated videos, identifying patterns and commonalities that could help us filter them out. Based on this analysis, we removed videos containing specific keywords and hashtags associated with irrelevant content.

In the second step, we used Google Gemini\footnote{\url{https://gemini.google.com/}\label{note3}} to classify whether a video was related to eating disorders. Gemini analyzed video descriptions to assess their relevance to eating disorder content. We first applied a classification prompt to a random sample of 200 videos (see Table~\ref{table:prompt} for the prompt). To evaluate its performance, we manually reviewed the results and found that Gemini achieved approximately 99\% accuracy. Based on this performance, we applied the same prompt to the full dataset.

\begin{figure*}[h]
  \includegraphics[width=.95\textwidth]{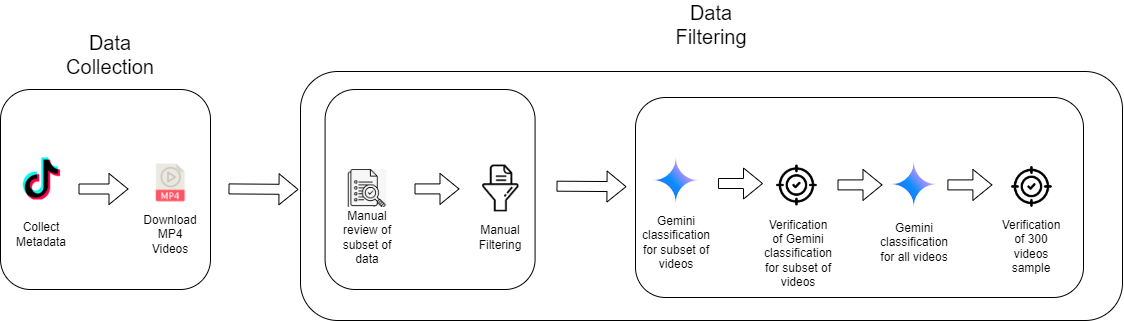}
  \caption{Dataset Flowchart}
  \label{fig:data}
\end{figure*}

To further verify the accuracy of the filtering process, we randomly sampled 300 videos from the filtered data, and all were confirmed to be related to eating disorders. After completing both processes, we finalized a dataset containing 43,040 videos. See Figure \ref{fig:data} for the full flowchart. Once we finalized the dataset we then collected the 388,332 comments and 188,739 replies for the videos, totaling 577,071 comments.

\section{Analysis}
In this section, we present an in-depth analysis of the dataset, focusing on engagement metrics as well as temporal, textual, and multimodal analyses. These analysis provide an overview of the dataset’s structure and thematic content, offering valuable insights for future research.

\subsection{Engagement Statistics}
We analyzed the engagement metrics for each video in our dataset, which included the number of likes, comments, shares, and views. The dataset comprises a total of 10,897 unique users, with engagement levels of: 79,876,956 likes, 537,329,790 views, 577,071  comments, and 962,040 shares (see Table \ref{table:stats}).

\begin{table}[h!]
\caption{Statistics on the dataset}
\begin{tabular}{|l|l|}
\hline
\textbf{Attribute} & \textbf{Sum} \\ \hline
No. Users          & 10,897        \\ \hline
No. Likes          & 79,876,956   \\ \hline
No. Views          & 537,329,790  \\ \hline
No. Comments       & 577,071      \\ \hline
No. Shares         & 962,040       \\ \hline
\end{tabular}
\label{table:stats}
\end{table}

\subsection{Temporal Analysis}
In this section, we analyze the temporal trends in the dataset. These analyses provide insights into the changing volume of content related to eating disorders and its impact on user interaction over the dataset's time span. 

\begin{figure*}[h]
    \centering
    \includegraphics[width=.9\textwidth]{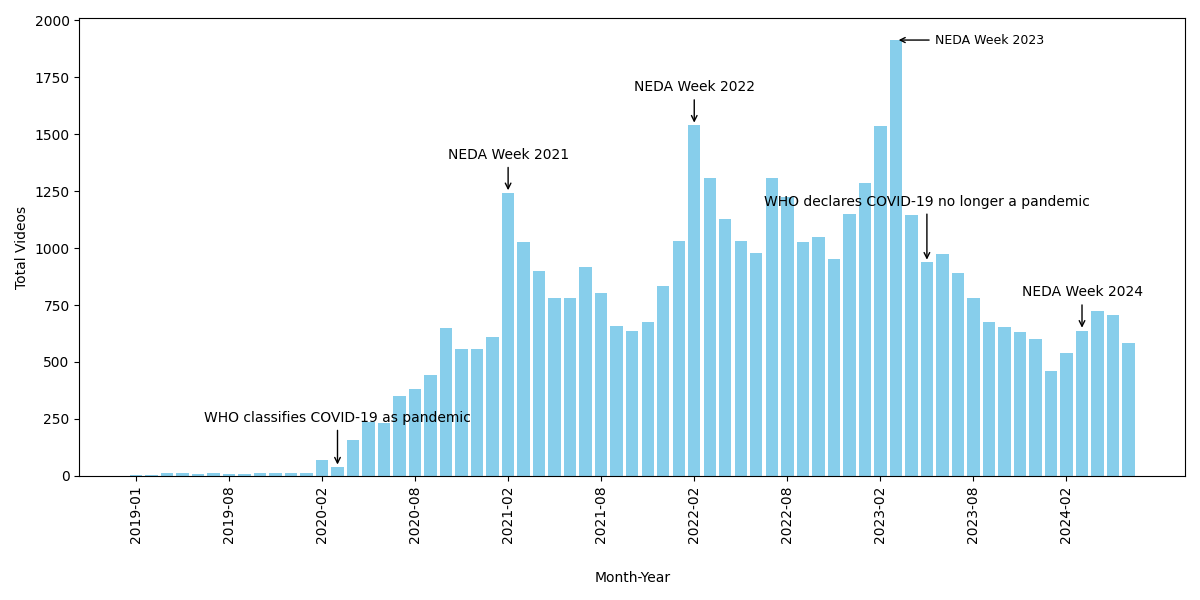}
    \caption{Total occurrences of videos per month. WHO: World Health Organization, NEDA: National Eating Disorder Awareness}
    \label{fig:videos_per_month}
\end{figure*}

Figure \ref{fig:videos_per_month} shows the number of videos posted per month across the span of our dataset. The graph reveals a noticeable growth in eating disorder-related videos after March 2020, the month COVID-19 was declared a global pandemic. There are strong seasonal patterns through 2023: peaks observed in February and March are likely due to National Eating Disorder Awareness Week sponsored by the National Eating Disorders Association (NEDA). Manual inspection of videos confirms that many videos posted during this time focus on raising awareness about eating disorders and educating the public about this deadly condition. Starting  mid-2023, there is a gradual decline in the volume of eating disorder-related content. While we can only speculate about the reasons for this decline, it could be attributed to TikTok’s efforts to moderate content related to eating disorders.

\begin{figure*}[t]
    \centering
    \includegraphics[width=.9\textwidth]{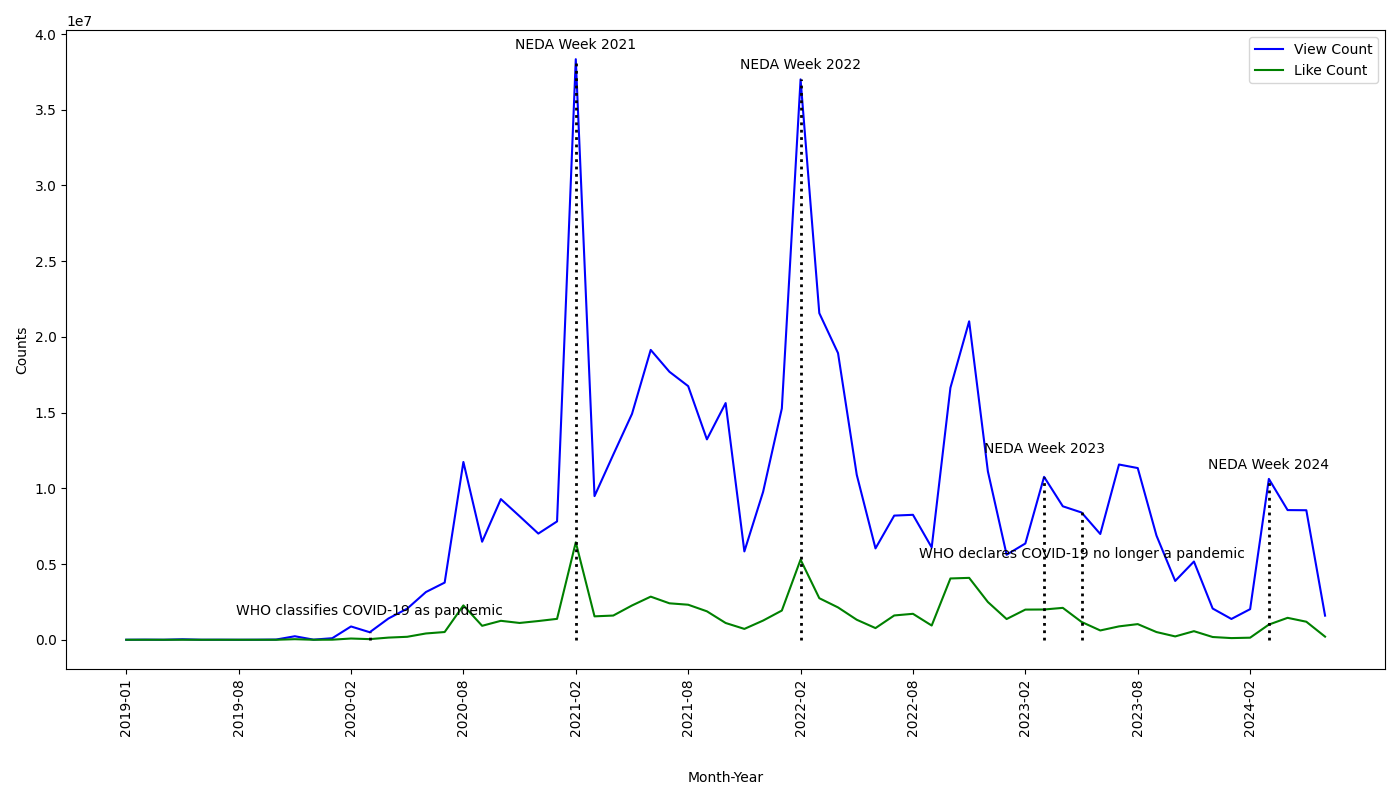}
    \caption{View and Like Counts of videos per month. WHO: World Health Organization, NEDA: National Eating Disorder Awareness}
    \label{fig:engagement_per_month}
\end{figure*}

Figure \ref{fig:engagement_per_month} shows the engagement metrics---views and likes count---throughout the period of our dataset. The view count (blue line) exhibits significant fluctuations, with several sharp peaks, particularly around mid-2020 and early 2021. This again could correspond to a rise in public attention to eating disorders during these periods, potentially influenced by major events like the COVID-19 pandemic, which saw an uptick in social media usage, as well as National Eating Disorder Awareness Week. The like count (green line) shows a more gradual trend, with a noticeable increase during the same periods, though at a much lower scale compared to views. Like counts follow the same general trend as views but remain less volatile. 

The engagement drops in 2023 may suggest TikTok’s moderation efforts targeting content related to eating disorders, as fewer videos and interactions are captured in later periods. This decline could also reflect shifts in user behavior or changes in content visibility on the platform.

\subsection{Text Analysis}
In this section we describe the textual analysis performed on our dataset. We used both the video descriptions and comments to do these analysis.

\begin{figure*}[t]
    \centering
    \includegraphics[width=.9\textwidth]{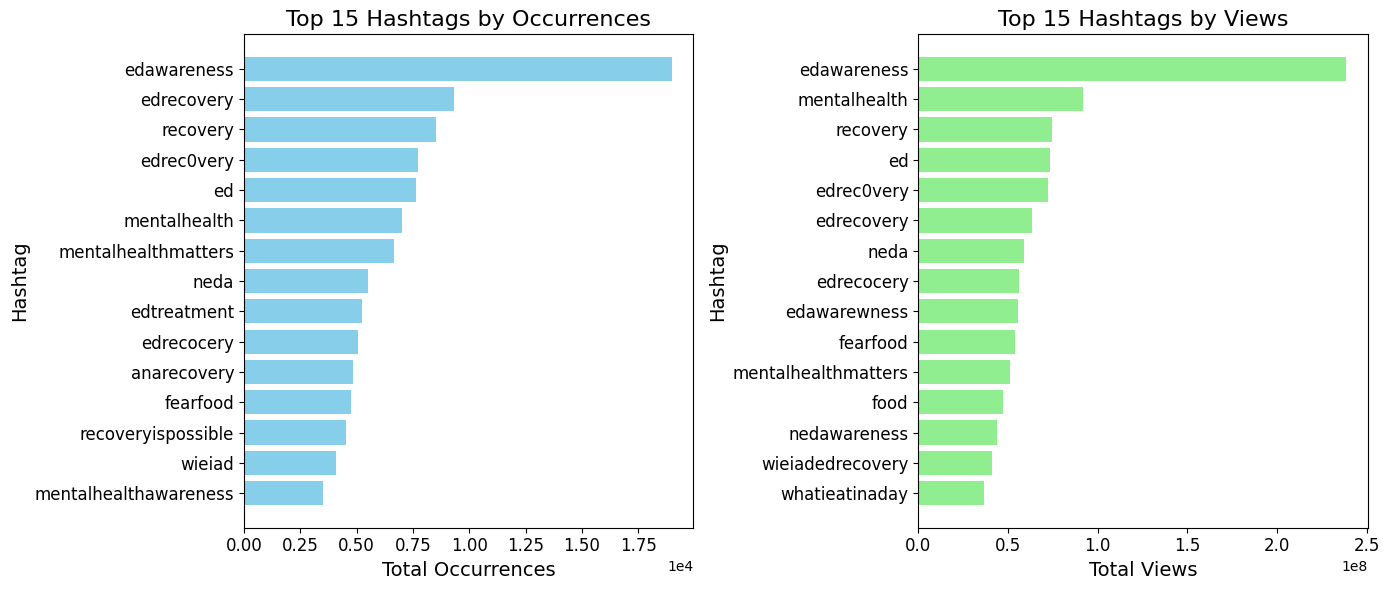}
    \caption{Top 15 Most Frequent and Viewed Hashtags}
    \label{fig:top15}
\end{figure*}

Figure \ref{fig:top15} shows the top 15 most popular hashtags in the dataset, i.e., most frequently used and most viewed. The most frequently used hashtag, ``\#edawareness,'' stands out with a significantly higher count than the others, emphasizing the focus on raising awareness about eating disorders. Hashtags like ``\#edrecovery,'' ``\#recovery,'' and ``\#edrec0very'' follow closely, reflecting the community's emphasis on recovery and support for individuals dealing with eating disorders. Other prominent hashtags such as ``\#mentalhealth,'' ``\#mentalhealthmatters,'' and ``\#neda'' (National Eating Disorders Association) suggest that content often intersects with broader mental health issues. Interestingly, hashtags like ``\#fearfood'' and ``\#recoveryispossible'' indicate content focused on specific challenges and motivational aspects of eating disorder recovery. The prominence of awareness and recovery-focused hashtags aligns with broader social media trends, where users seek both to spread knowledge about eating disorders and to foster supportive communities. These findings highlight how TikTok users use specific hashtags to categorize their content, with a strong emphasis on recovery and awareness efforts.

Figure \ref{fig:cloud} shows a word cloud for the video descriptions provides insights into the most frequently used words in the dataset after removing hashtags and stopwords. By tokenizing the words and counting their frequency, several key terms emerge, reflecting common themes in the content. The most frequent word, ``ed'' (eating disorder), appears 1,544 times, underscoring the central focus of the videos. Other prominent words like ``today'' (1,474), ``day'' (1,397), and ``recovery'' (1,308) suggest that many videos emphasize daily struggles and progress within the context of recovery. Words like ``eat'' (1,290), ``food'' (1,251), and ``fear food'' (649) highlight the central role food plays in the discourse, often framed in terms of challenge and fear. Emotional and supportive language is also common, as seen in terms like ``love'' (1,160), ``feel'' (706), and ``good'' (705), reflecting the users' engagement with self-perception, emotion, and mental well-being. Additionally, the appearance of ``people'' (703) and ``help'' (622) suggests a focus on community and support. This word cloud highlights the key areas of discussion within the video descriptions, which center around recovery, daily experiences, food challenges, and emotional well-being.


Using BERTopic \cite{grootendorst2022bertopic} to analyze the video descriptions allows us to uncover distinct themes or topics based on the textual data. BERTopic leverages transformers like BERT, which captures contextual embeddings of words, and a clustering algorithm group similar documents. The result is a set of interpretable topics representing the primary themes of the data.

We applied topic modeling to the video descriptions and then gave them certain themes based off the keywords for the topics see Table \ref{tab:topics_descriptions}. Weight/Body Image is a recurring theme, with words like ``weight,'' ``skinny,'' and ``gain,'' indicating discussions around weight-related concerns, body image, and possibly weight loss or gain journeys. Fear/Anxiety also surfaces, characterized by terms such as ``scared,'' ``afraid,'' and ``terrified,'' suggesting prevalent discussions about managing fears and anxiety that co-exist in those struggling with eating disorders. Similarly, Guilt/Shame appears frequently, with mentions of ``guilt,'' ``shame,'' and ``regret,'' reflecting the strong emotional experiences accompanying this condition. The Crying/Emotions topic stands out, where words like ``cry,'' ``tears,'' and ``sobbed'' reflect expressions of deep emotional release or sadness. The topic Treatment/Healthcare uses terms such as ``inpatient,'' ``hospital,'' and ``residential'' points to conversations around medical treatments, support structures, and rehabilitation programs for those struggling with eating disorders. The topic of ``Challenges'' emerges, with terms like ``challenge'' and ``challenging'' often linked to viral trends or users pushing themselves to overcome personal obstacles. Eating Disorders emerge prominently, with words such as ``anorexia'' and ``disorder'' signaling discussions centered on the struggles with specific conditions and their impacts. The Eating theme is also significant, characterized by mentions of ``meal,'' ``eat,'' and ``plan,'' reflecting ongoing efforts to establish healthy eating habits or navigate meal planning during recovery. The focus on Eating Disorder Recovery emerges with terms like ``recovery'' and ``reclamation,'' underscoring the continuous journey toward healing and finding a balanced relationship with food. Lastly, the Hunger theme, with words like ``hungry'' ``ate,'' and ``extreme,'' points to feelings of being overwhelmed by hunger or dealing with intense cravings, potentially as a part of disordered eating behaviors or recovery struggles.
\begin{table}[h]
    \centering
    \caption{Video Description Topics and Words Associated}
    \begin{tabular}{@{}ll@{}}
        \toprule
        \textbf{Topic}                & \textbf{Top Words}                  \\ 
        \midrule
        Weight/Body Image                & weight, skinny, gain            \\ 
        \hline
        Fear/Anxiety           & scared, afraid, terrified                \\ 
        \hline
       Guilt/Shame                & guilt, shame, regret                  \\ 
        \hline
        Crying/Emotions               & cry, tears, sobbed          \\ 
        \hline
        Treatment/Healthcare                    & inpatient, hospital, residental             \\ 
        \hline
        Challenges                & challenge, challenging                \\ 
        \hline
        Eating Disorders                & anorexia, disorder              \\ 
        \hline
        Eating             & meal, eat, plan       \\ 
        \hline
        ED recovery               & recovery, reclamation             \\ 
        \hline
        Hunger                    & hungry,ate, extreme                  \\ 
        \bottomrule
    \end{tabular}
    \label{tab:topics_descriptions}
\end{table}

We also applied topic modeling to analyze the comments and replies we collected, revealing several significant themes see Table \ref{tab:topics_comments}. The Recovery theme features words like ``recovery,'' ``recovering,'' and ``choosing,'' indicating discussions centered on the healing process, the journey toward full recovery, and the choices individuals face along the way. The Eating Disorders (ED) topic emerges strongly, with terms like ``ed'' and ``dysfunction'' reflecting conversations about the challenges and treatment associated with eating disorders. The Battle Fighting metaphor emerged, evident in words such as ``fighting,'' ``battle,'' and ``fighter,'' which frame recovery as an ongoing battle against adversity. The Pride theme emerges with terms like ``proud'' indicating supportive and encouraging comments, often directed at others. The Gratitude topic emphasizes expressions of appreciation, featuring words like ``thank,'' ``muchhh,'' and ``genuinely,'' indicating the importance of community support and acknowledgment in the recovery process. Discussions around Hunger are also prominent, with terms like ``hunger,'' ``extreme,'' and ``hungry,'' likely addressing the challenges of extreme hunger or cravings that individuals with eating disorders often experience. This could point to commenters empathizing the video creator by mentioning struggling with similar challenges. The Weight Gain topic includes words like ``gaining,'' ``weight,'' and ``pounds,'' signifying the complexities and emotional weight associated with weight gain throughout users' recovery journeys. The Hugs theme highlights the supportive nature of the commenters, featuring words like ``hug,'' ``give,'' and ``big,'' which reflect a desire for connection and comfort among community members. The Journey theme, characterized by terms like ``journey,'' ``following,'' and ``luck,'' underscores the communal aspects of recovery, emphasizing shared experiences and mutual support. Lastly, the Struggling theme is marked by words such as ``struggle'' and ``suffer,'' revealing the persistent challenges users face throughout their recovery journeys.

\begin{table}[h]
    \centering
    \caption{Comments and Replies Topics and Words Associated}
    \begin{tabular}{@{}ll@{}}
        \toprule
        \textbf{Topic}                & \textbf{Top Words}                  \\ 
        \midrule
        Recovery                & recovery, recovering, choosing            \\ 
        \hline
        Eating Disorders           & ed, dysfunction              \\ 
        \hline
       Fighting Battle                & fighting, battle and fighter                \\ 
        \hline
        Pride              & so, proud         \\ 
        \hline
        Gratitude                    & thank, muchhh, genuinely             \\ 
        \hline
        Hunger                & hunger, extreme, cues, appetite             \\ 
        \hline
        Weight Gain                & gaining, weight, pounds              \\ 
        \hline
        Hugs                & hug, give, big           \\ 
        \hline
        Journey                & journey, following, luck              \\ 
        \hline
        Struggling                  & struggle, suffer                 \\ 
        \bottomrule
    \end{tabular}
    \label{tab:topics_comments}
\end{table}

\begin{figure*}[t]
    \centering
    \includegraphics[width=.9\textwidth]{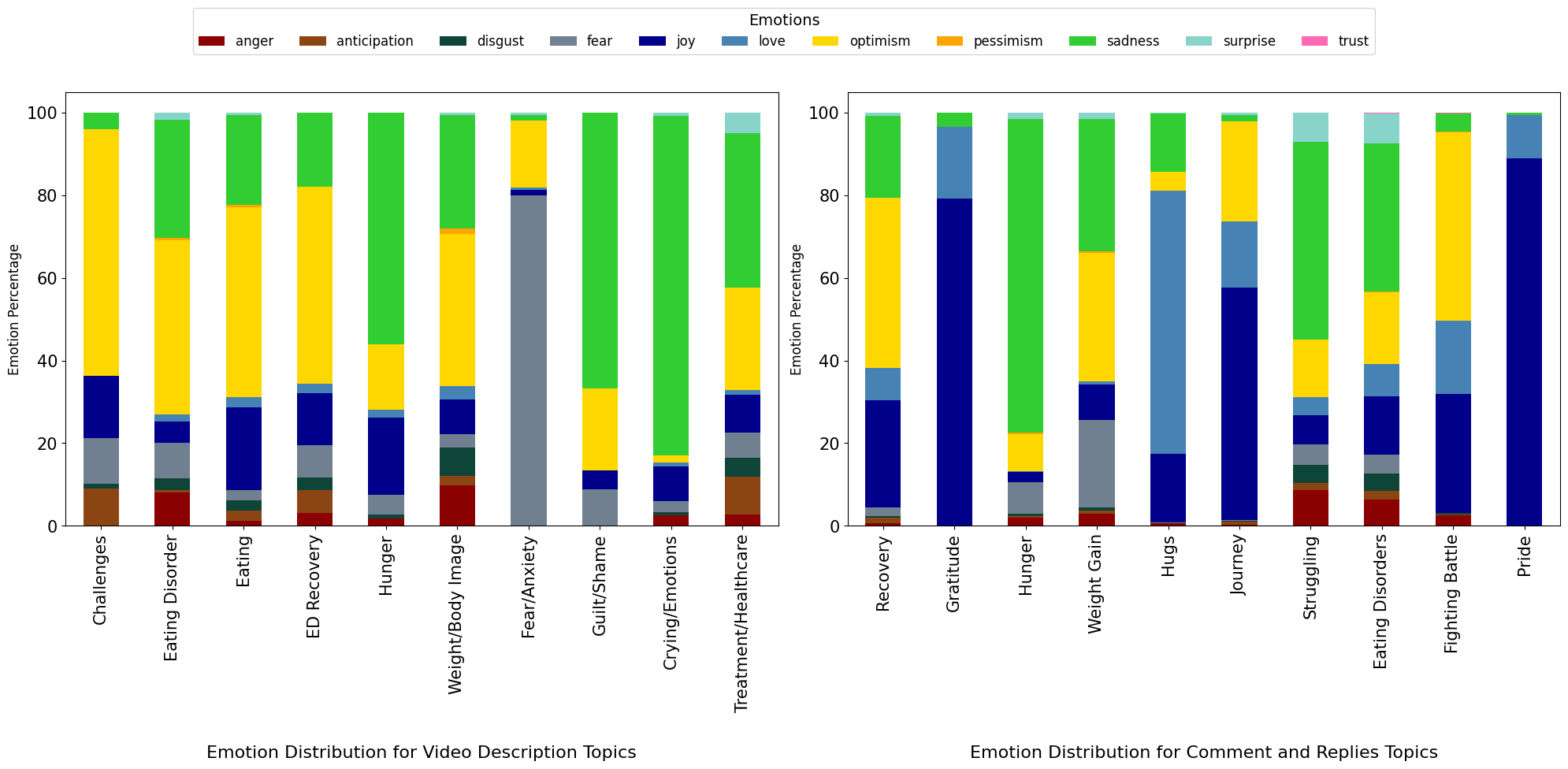}
    \caption{Comparative emotion analysis of video descriptions and comments and replies topics}
    \label{fig:emotion_distrubution}
\end{figure*}

We followed the topic modeling by doing an emotional analysis per topic. We utilize an emotion detection tool called Demultiplexer (Demux) \cite{chochlakis2023leveraging}, inspired by SpanEmo \cite{alhuzali2021spanemo}. The tool processes two input sequences: the first sequence represents the emotion categories, and the second sequence contains the actual content. Demux leverages contextual embeddings for each emotion to calculate the probabilities of specific emotional expressions. We applied Demux to analyze every video description and comment, focusing on emotions such as anger, anticipation, disgust, fear, joy, love, optimism, pessimism, sadness, surprise, trust and excluding the text that were classified as None. We then did an emotional analysis for each topic for the video descriptions and comments see Figure \ref{fig:emotion_distrubution}.

The emotional analysis of the video descriptions reveals insights in how various emotions manifest across TikTok videos regarding eating disorders. The Weight/Body Image topic reflects significant levels of anger (9.77\%) and sadness (27.36\%), suggesting that concerns about body image elicit intense emotional responses, as individuals struggle with feelings of frustration and despair about their appearance. However, optimism (36.81\%) also emerges, indicating that, despite these negative emotions, users maintain a hopeful outlook, possibly reflecting aspirations for self-acceptance or improvement. The Fear/Anxiety theme is overwhelmingly associated with fear (80.00\%), revealing a profound sense of anxiety experienced by individuals concerning their eating habits and body image, which highlights the emotional toll these issues take on their well-being. The Guilt/Shame theme also reveals a significant degree of sadness (66.67\%), emphasizing the heavy emotional burden often felt in relation to eating behaviors and self-perception. Similarly, the Crying/Emotions topic underscores this emotional distress, with a striking sadness level (82.20\%), indicating that individuals frequently express their feelings through tears and emotional releases linked to their struggles. In the context of Treatment/Healthcare, the emotional landscape is characterized by sadness (37.29\%) and optimism (24.86\%), reflecting both the emotional challenges associated with seeking help and the hope for positive change that often accompanies treatment experiences. Moving to the Challenges theme, we observe a signficant amount of optimism (59.59\%) this may indicate that users see challenges as opportunities for growth, self-improvement, or community support. The Eating Disorder topic showcases significant sadness (28.57\%) and optimism (42.29\%), emphasizing the deep emotional struggles tied to these disorders while simultaneously revealing a sense of hope for recovery. The Eating theme indicates a balanced emotional landscape with joy (19.88\%) and optimism (45.96\%), suggesting that discussions surrounding food can elicit positive feelings alongside the complexities of eating habits. Finally, ED Recovery demonstrates a notable level of optimism (47.66\%) alongside some sadness (17.97\%), highlighting the hopefulness surrounding recovery processes, even amid emotional challenges. The Hunger topic stands out with a high level of sadness (56.07\%), reflecting the emotional weight of hunger and its association with disordered eating behaviors. This emotional breakdown offers valuable insights into the varied emotional responses within these discussions, illustrating how some topics evoke feelings of fear and sadness while others foster hope and optimism.

The emotional analysis of the comment and replies topics reveals distinct emotional patterns within TikTok discussions surrounding eating disorders and recovery experiences. The Recovery theme indicates a notable presence of joy (26.04\%) and optimism (41.21\%), suggesting that many individuals find hope and positivity in their healing journeys, despite also experiencing sadness (19.79\%) and fear (2.01\%). The Gratitude topic stands out for its overwhelming expression of joy (79.26\%), highlighting the importance of thankfulness and appreciation within the community, with minimal representation of other emotions. In contrast, the Hunger theme shows a strikingly high level of sadness (75.90\%), reflecting the emotional turmoil associated with feelings of hunger and disordered eating behaviors. The Weight Gain discussions reveal significant levels of fear (21.12\%) and sadness (32.06\%), indicating the anxiety related to weight fluctuations and the emotional complexities tied to body image. The Hugs topic emphasizes emotional support, characterized by a high level of love (63.79\%), suggesting a nurturing environment among community members. The Journey theme reveals a prominent joy (56.19\%), indicating a sense of positivity associated with personal growth and experiences shared within the community. Conversely, the Struggling topic exhibits high levels of anger (8.67\%) and sadness (47.78\%), highlighting the significant emotional distress and challenges faced by individuals navigating their journeys. The Eating Disorders topic reflects a mix of sadness (35.84\%) and joy (14.01\%), underscoring the emotional toll of these disorders while also recognizing moments of happiness. Lastly, the Fighting Battle theme reveals a strong sense of optimism (45.60\%) alongside joy (28.86\%), indicating that while recovery is framed as a struggle, there is also a prevailing hopefulness among individuals facing these challenges. This emotional breakdown highlights the complex emotional landscape surrounding eating disorders, with some discussions fostering fear and sadness, while others promote joy, love, and optimism.

\subsection{Multimodal Analysis}
For the multimodal analysis, we randomly sampled over 500 videos along with their video descriptions. Using the Gemini model, we generated detailed descriptions of each video (see Table~\ref{table:prompt_video} for the prompt). The outputs offered valuable insights into video themes and content, which we systematically analyzed to identify recurring patterns. This analysis revealed four prominent themes that consistently appear in discussions related to eating disorders, body image, and food anxieties.

A significant number of videos center around the concept of ``fear foods,'' where creators openly confront foods they have previously avoided due to anxiety or fears about weight gain. Ability to control anxiety and eat these foods is a significant part of the recovery process. These videos aim to document the creators' recovery journeys and create a sense of community among viewers facing similar struggles.. 

Another major theme present in the videos is the critique of diet culture and societal beauty standards. Many creators use their platform to engage in anti-anorexia speech, challenge restrictive dieting practices, calorie counting, and the glorification of thinness. These videos encourage viewers to reassess societal expectations regarding the body  and adopt a more balanced approach to their relationship with food. By questioning the widespread promotion of ``ideal'' body types, these creators aim to dismantle the harmful effects of diet culture and advocate for a more inclusive definition of beauty.

The promotion of body positivity and body neutrality emerges as a third key theme, as creators seek to encourage the acceptance of diverse body types and normalize natural body features such as bloating or weight fluctuations. These videos often challenge unrealistic beauty standards by providing a counter-narrative that emphasizes self-acceptance and the idea that a person’s worth is not tied to their appearance. Some creators extend this message to a body-neutral approach, aiming to shift focus away from appearance altogether and prioritize mental and physical well-being.

Lastly, the sharing of personal recovery narratives is a powerful and recurring theme in the videos. Many creators openly discuss their experiences with eating disorders, self-harm, and the ongoing journey towards recovery. These narratives often serve to offer hope, inspire resilience, and foster a sense of shared experience, creating a supportive space where viewers can find validation for their struggles. While these personal stories often have a positive impact, there is also a need to remain mindful of the potential for content to be triggering for some individuals, particularly when discussing sensitive topics such as weight, diet, and body image.

Together, these themes illustrate how TikTok has become a critical space for addressing eating disorders and body image concerns, highlighting the platform’s dual role in both challenging harmful cultural norms and potentially perpetuating them. This emphasizes the importance of careful content creation and consumption within online communities dealing with mental health and body-related issues.

\section{Discussion}

This dataset offers a unique opportunity for examining the landscape of eating disorders as it pertains to social media, with a specific focus on TikTok. This dataset includes not only metadata but also user-generated comments and the mp4 video files associated with each post. The multimodality allows for a more comprehensive analysis of how eating disorder-related content is created, shared, and consumed on TikTok. Furthermore, this dataset serves vulnerable populations, particularly adolescents and young adults, who are disproportionately affected by eating disorders and spend a significant amount of time on social media platforms. By capturing the multimodal nature of social media interactions, this dataset is well-suited for investigating the complexities of eating disorder content, providing insights that go beyond text-only analyses.

The importance of this dataset extends beyond its unique composition; it addresses a significant societal issue. Eating disorders are a critical public health concern, with millions of adolescents and young adults experiencing these disorders worldwide. Social media platforms, including TikTok, have emerged as influential environments where young people discuss body image, dieting, and recovery. Therefore, analyzing how eating disorder content is presented on TikTok, including the emotional tone conveyed in video descriptions and user interactions, can shed light on potential harms and benefits associated with such content. Understanding these dynamics is particularly important for vulnerable populations who may be at a higher risk of engaging with harmful content. Insights derived from this dataset could inform public health interventions, content moderation policies, and mental health support systems on social media.

The data also opens up various research questions that could guide future studies in this area. For instance, the pandemic has led to increased social isolation, changes in daily routines, and elevated stress levels, all of which are factors that may exacerbate eating disorders. A longitudinal analysis of the dataset could help in understanding the changes in discourse related to eating disorders before, during, and after the pandemic. Additionally, the dataset could be used to study the effectiveness and gaps in TikTok’s content moderation strategies regarding eating disorder-related content. Are harmful videos being flagged or removed, and if so, are there instances of beneficial content—such as recovery-focused discussions—being mistakenly censored? This line of inquiry could provide valuable insights into improving content moderation policies to better protect vulnerable users while promoting supportive communities.

Furthermore, the multimodality capability offers the potential to analyze how visual content, such as video aesthetics, aligns or contrasts with the textual sentiment expressed in captions and comments. This could help understand the role of visual media in conveying certain messages about body image and eating behaviors, thus offering a holistic view of the content and its potential impact.

Despite these strengths, several limitations must be acknowledged. Our dataset was collected using a predefined set of hashtags and keywords, supplemented by manual curation, which may introduce bias by failing to capture the full spectrum of eating disorder-related content on TikTok. Additionally, classification was conducted using Google Gemini without model tuning, which could affect filtering accuracy. Lastly, TikTok's Research API imposes constraints such as rate limits, token expiration, and daily request caps, making large-scale data collection time-intensive. 

\section{Ethics}
In compliance with the TikTok Research API Terms of Service, we regularly updated and pruned our dataset to ensure that it remained aligned with TikTok’s current API guidelines. No identifiable information is included in the dataset to protect the privacy of TikTok users, and no attempts were made to identify individual users or access private data. This research was conducted exclusively for academic purposes, with the primary goal of advancing the understanding of social media’s relationship to eating disorders. We were mindful of the sensitive nature of topics like eating disorders and approached the content with caution and respect. The research was designed to contribute meaningfully discussions on social media and public health without exploiting or sensationalizing individual experiences.

\section{Data Availability}
The dataset, accessible at 
\href{https://github.com/cbickham3232/EDTok-A-Dataset-for-Eating-Disorder-Content-on-TikTok}{\nolinkurl{https://github.com/cbickham3232/EDTok-A-Dataset-for-Eating-Disorder-Content-on-TikTok}}, adheres to TikTok's Terms and Conditions, which restrict the public distribution of collected videos. As an alternative, we provide the Video IDs—unique identifiers corresponding to specific posts. Researchers can use these IDs to query TikTok’s API and obtain the complete video objects, including both multimedia content and metadata.

\section{Conclusion}
In conclusion, this dataset serves as a valuable resource for understanding the multifaceted nature of eating disorder-related content on TikTok, especially among adolescents and young adults. Its unique structure enables the exploration of various aspects of the content, offering a deeper understanding of the issues at hand. Given the significant impact of eating disorders on vulnerable populations, leveraging this dataset for research can contribute to more informed content moderation policies, better mental health support strategies, and targeted public health interventions.

\bibliography{aaai22}
\appendix
\begin{table}[]
\caption{Prompt input template for detecting if a video is related to eating disorders}
\begin{tabular}{|p{0.85\columnwidth}|} 
\hline
\textit{\begin{tabular}[c]{@{}p{0.8\columnwidth}@{}} 
Task: You are an expert in psychiatry and social media. You are tasked with classifying whether a video description is related to eating disorders based on the content of the description.\\
For each id and video description pair, return a JSON object where the key is the id and the value is 1 if the description indicates a video about eating disorders and 0 if it does not.\\
Criteria for eating disorder classifications:
\begin{itemize}
    \item[-] Mentions of terms related to eating disorders such as ``anorexia,'' ``bulimia,'' ``binge eating,'' ``dieting,'' ``fasting,'' ``weight loss,'' ``body image,'' ``calories,'' or any other phrases related to disordered eating habits or behaviors
    \item[-] Implicit descriptions of unhealthy eating patterns, extreme dieting practices, or body dysmorphia
    \item[-] Videos promoting or depicting recovery from eating disorders, including terms like ``recovery,'' ``ED recovery,'' or ``mental health support'' related to eating behaviors
\end{itemize}
Input format:
List of video id and video descriptions:\\
{[}id 1: video description{]}\\
{[}id 2: video description{]}\\
.........\\
{[}id 100: video description{]}
\end{tabular}} \\ \hline
\end{tabular}
\label{table:prompt}
\end{table}

\begin{figure*}[t]
    \centering
    \includegraphics[width=0.7\textwidth]{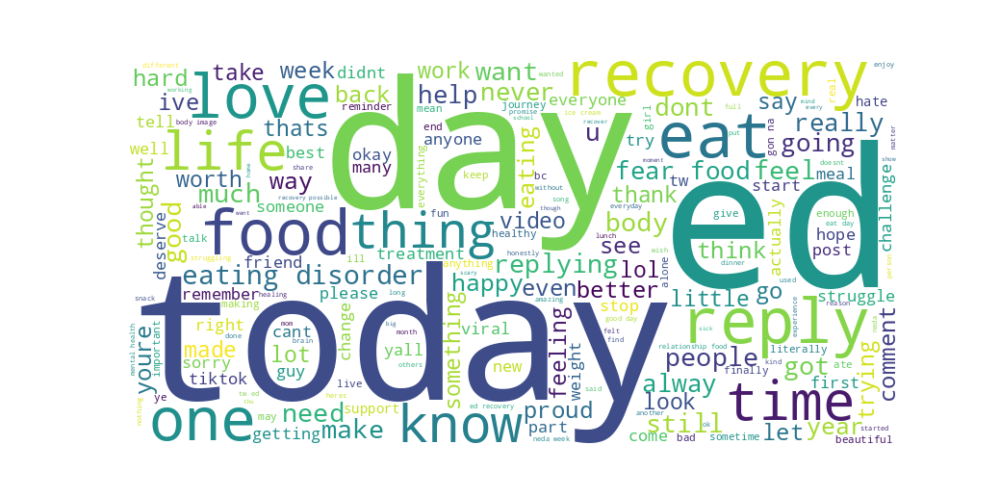} 
    \caption{Word Cloud of Video Descriptions}
    \label{fig:cloud}
\end{figure*}

\begin{table}[]
\caption{Prompt input template for analyzing videos and comments}
\begin{tabular}{|p{0.85\columnwidth}|} 
\hline
\textit{You are an expert in psychiatry and social media. Please analyze the following TikTok video and its video description with a focus on identifying key psychological and social themes.} \\ \hline
\end{tabular}
\label{table:prompt_video}
\end{table}
\end{document}